\documentclass[doublecol,linenumbers]{epl2} 
\usepackage{bbold}
\usepackage{braket}
\usepackage{epstopdf}
\usepackage{amsmath,amssymb,mathtools}

\title{Representations of tensor rotations and the geometry of spin 1/2}
\shorttitle{Representations of tensor rotations and the geometry of spin 1/2} 

\author{M. B\"uhler}
\shortauthor{M. B\"uhler}

\institute{                    
  \inst{1} First Institute - Address\\
  \inst{2} Second Institute - Address
}

\pacs{03.65.Sq}{Semiclassical theories and applications}
\pacs{04.30.-w}{Gravitational waves}
\pacs{03.50.Kk}{Other classical field theories}

\abstract{
Making use of the real sl(2,$\mathbb{R} $) Lie group algebra generating a spin 1/2 Lie group allows to create an explicitly given Lorentz invariant fermion wave. As the generators are real valued they can be interpreted as a deformation tensor in particular as a deformation tensor of space. Therefore, it is possible to model a heuristic purely geometric representation of spin 1/2 in Minkowski space. However the bigger surprise is that this wave has the space-time structure of gravitational waves, which are understood to be spin 2 waves. Given that the uniqueness of angular momentum representations still holds, the examination of tensor rotations reveals the existence of different representations of tensor rotations with a different angular parameter due to an unaccounted basic symmetry of symmetric tensors, where the spin 1/2 representation is a specific representation of tensor rotations corresponding to the quantum theoretical approach. The seeming contradiction is fully resolved and allows in addition to understand the notion of different representations of spin in tensors, again related to different representations of the tensor. 
}

\begin{document}

\maketitle

\section{Introduction}

So far there is no 
representation of spin 1/2 in Euclidean space known that could help to give a more comprehensible understanding of this crucial feature of quantum mechanics. However, there have been successful attempts to rewrite spin in terms of geometric algebra, mainly by Hestenes and others \cite{hestenes1985cliffordalgebra,Sobczyk:2015txa } and attempts to understand the double cover provided by spin 1/2 in real space \cite{misner1973gravitation}.
Total angular momentum conservation of quantum mechanics shows that the sum of orbital angular momentum and any internal or spin angular momentum has to be included in the conservation of angular momentum and that orbital angular momentum can be transferred to spin angular momentum and vice versa. Therefore one would expect that both angular momenta have something in common, which are the common angular momentum commutator relations of quantum mechanics. In the classical case, we know that orbital angular momentum is related to the rotational invariance and equivalently to a coordinate system rotation but a similar notion doesn't exist of spin, mainly because the internal space is seen as an abstract complex space reflecting the internal degrees of freedom of particles. Therefore the problem of having no notion of spin might be related to the missing notion of the internal space.

\section{Constructing a spin 1/2 geometric model}
\begin{figure}
\centering\includegraphics[width=3.4in]{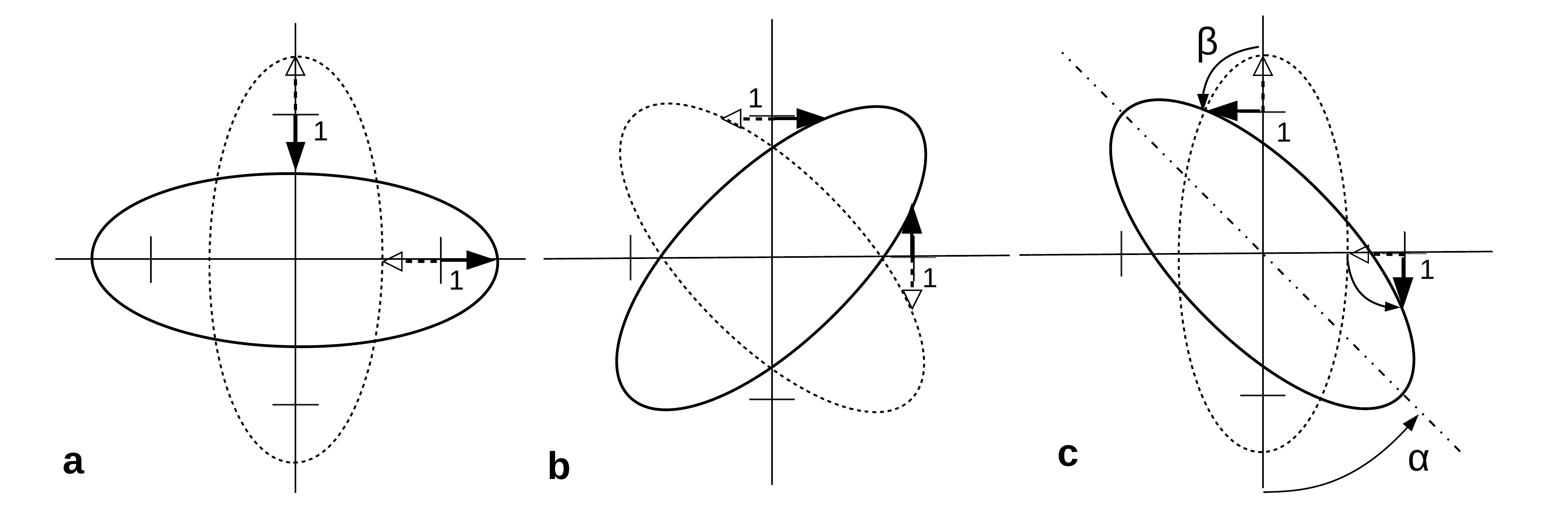}

\caption{A geometric representation of the action of the generators of SL(2,$\mathbb{R}$) in the plane perpendicular to the propagation. Small arrows give the components of the matrix. a) The linear polarized wave created by $g_1$. The wave oscillates between the full and dotted line ellipse by stretching in one coordinate direction and squeezing in the perpendicular direction while propagating. b) The linear polarized wave by $g_2$. Same oscillation as in a) but stretching/squeezing in the diagonal directions. c) Generator $g_3$ rotates the ``internal vectors'' represented by the arrows. A $\beta=4\pi$ rotation is necessary to rotate the ellipse by $2\pi$ as required for a spin 1/2 excitation. The representation of the rotation is not unique as the same rotation can be done by rotating the ``internal vectors'' by $\beta=4\pi$ or by rotating the major and minor axes with $\alpha=2\pi$ as indicated.}

\label{fig:gen3}
\end{figure}

In order to artificially construct a purely geometric model of spin 1/2 we use the real sl(2,$\mathbb{R} $) Lie group algebra and apply it to the Euclidean space, which is seen here as an elastic medium \cite{tiffen1970plane}. The purpose of using sl(2,$\mathbb{R} $) is twofold: the commutator relations as summarized in eq.~\eqref{eq:sl2r} qualify it to be a spin 1/2 representation\cite{feynman1998statistical} and allows for a Poincar\'{e} invariant geometric description. In order to have a spin operator as in quantum mechanics on would multiply the generators used here with $1/2\hslash$ similar to the Pauli matrices where $\hslash$ is the reduced Planck constant. In our case the generators are real valued and therefore the application of the generators can be interpreted as an elastic deformation, in particular as a deformation of Euclidean space. We consider infinitesimal small changes only keeping the Euclidean space to an arbitrary good approximation. 
\begin{equation}\label{eq:sl2r}
 \begin{split}
 & \tx{generic generators of SL(2,$\mathbb{R}$):}\\
 &H= \begin{pmatrix} 1&0\\0&-1 \end{pmatrix},~X=\begin{pmatrix} 0&1\\0&0 \end{pmatrix},~Y=\left(\begin{matrix} 0&0\\1&0 \end{matrix}\right) \\
 &~X=a,~Y=a^\dag,~ N=a a^\dag=\left(\begin{matrix} 1&0\\0&0 \end{matrix}\right)\\
 &\tx{fermion commutator relations:}\\ 
 &\{a^\dag,a\}=1,~ \{a^\dag,a^\dag\}=\{a,a\}=0\\ 
 & [N,a^\dag ]=a^\dag,~ [N,a]=-a
  \end{split}
\end{equation}
Because of the 2-dimensional matrix representation of the sl(2,$\mathbb{R} $) algebra we are going to use in order to allow a spin 1/2 representation, we assume a harmonic wave propagating in $x_3$ direction with the speed of light in vacuum, which restricts the polarizations to be transversal with two polarizations \cite{wigner1939ireps,cesare2017fieldapprtogw} and allows to adapt the dimensions of Euclidean space and SL(2,$\mathbb{R} $) by using the 2-dimensional subspace perpendicular to the propagation as in eq.~\eqref{eq:gbasis}. As shown below one arrives at the same irreducible representation asking for a light like wave in Wigners\cite{wigner1939ireps} classification, bringing us to the induced representations, the little group. But here the focus was on the possibility of a geometric interpretation.
\begin{equation}\label{eq:gbasis}
 \begin{split}
 &\tx{geometric basis:}\\
 &g_1=H=\begin{pmatrix} 1&0\\0&-1 \end{pmatrix},~g_2=a+a^\dag=\begin{pmatrix} 0&1\\1&0 \end{pmatrix},\\
 &g_3=a-a^\dag=\begin{pmatrix} 0&1\\-1&0 \end{pmatrix} \\
 &2D\mapsto3D:~\left( \begin{array} {cc} \tx{\small{SL(2,$\mathbb{R}$)}}&\begin{matrix} 0\\0\end{matrix}\\ \begin{matrix} 0&~0\end{matrix} &0 \end{array} \right)   \\
&\tx{model plane wave:}\\
 &\mathbb{1}+\begin{pmatrix}
 1   &0   &0  \\
 0   &-1  &0  \\
 0   &0   &0 
  \end{pmatrix}\tx{Re} \left(a_1\, e^{i\left(\omega t -k_3 x_3 \right)} \right) +\\
 & \qquad\qquad\qquad\begin{pmatrix}
 0  &1   &0  \\
 1  &0   &0 \\
 0  &0   &0 
  \end{pmatrix}\tx{Re} \left(a_2\, e^{i\left(\omega t -k_3 x_3\right)} \right) \\
 &\tx{Re: real part;} \left| a_i \right|\ll 1, \tx{$a_i$ complex;} \\
 & \tx{ $\omega$=~c~$k_3$, c: vacuum speed of light}
 \end{split}
\end{equation}
In order to allow a more geometric representation we introduce the new basis of generators in eq.\eqref{eq:gbasis} where $g_1$ and $g_2$ are the conjugacy class of mirror transformations creating the two polarizations and $g_3$ is the conjugacy class allowing rotations. The Lie group is generated by the matrix exponentiation of the generators \cite{hall2015Liegroups}, where we avoid the imaginary unit in the exponentiation as we work in real space only. In the limit of arbitrary small excitations the plane wave created is any linear combination of our generators $g_1$ and $g_2$ added to the identity matrix $\mathbb{1}$, which represents the undisturbed Euclidean space.
The geometric meaning of the wave is shown in figure~\ref{fig:gen3} where the deformation tensor in the transversal plane is plotted as an ellipse (There is no deformation of the ellipsoid in the propagation direction of the wave). In addition the matrix components of the generators responsible for the deformation are shown as small arrows added to basis vectors visualizing the mapping of the basis and corresponding to the columns of the matrix. In fig.~\ref{fig:gen3}a the result of applying the generator $g_1$ is shown as the full line ellipse. The wave oscillates between the full and the dotted line ellipse while propagating by stretching in one coordinate direction and squeezing in the perpendicular coordinate direction without changing the area of the ellipse (volume of the ellipsoid in 3 dimensions). The same applies to the wave generated by $g_2$ in fig.~\ref{fig:gen3}b now with the ellipse stretching/squeezing in the diagonal directions. Fig.~\ref{fig:gen3}c shows the application of the generator $g_3$ which rotates the ellipse. However this rotation of the ellipse is done by a rotation with angle $\beta$ of the arrow representing the vectors in the subspace of the SL(2,$\mathbb{R} $) group, not by a rotation with respect to the coordinates with angle $\alpha$, but is equivalent. Obviously we have the simple symmetry $\beta=2\alpha$. Mathematically this is due to the fact that in the two dimensional subspace where the rotation takes place with a given  symmetric two dimensional matrix $\tens{M}$ to be rotated it is always possible to separate into a static matrix proportional to the identity matrix $\mathbb{1}$ and a traceless matrix $\tens{V}$ as in eq.(\ref{eq:sep}) where $\tens{D}$ is the rotation matrix. The static matrix commutes with any other matrix and in particular with the rotation matrix. This decomposition is unique when the separated matrix $\tens{V}$ is chosen to be traceless and guaranties at the same time that the rotation applied to $\tens{V}$ can be represented by rotating vectors.
\begin{equation} \label{eq:sep}
 \tens{D} \tens{M} \tens{D}^{-1}= \tens{D}\left( \lambda\mathbb{1}+\tens{V} \right) \tens{D}^{-1}=\lambda\mathbb{1}+ \tens{D}\tens{V}\tens{D}^{-1} 
 \end{equation}
If we rotate the ellipse by referring to the ``internal coordinates'', given by the SL(2,$\mathbb{R}$) subspace, we need two full turns to rotate the ellipse by $2\pi$ which is exactly what is known for spin 1/2. Due to the symmetry of the tensor a single turn of the ``internal vectors'' gives the same deformation of the ellipse in accordance with quantum theory. Identifying the small elastic deformations with the ``internal space'' internal angular momentum conservation can be linked to the rotational invariance with respect to the ``internal coordinates''. Therefore spin, as modeled here, is related to ``internal rotations'' in the same way as orbital angular momentum is related to coordinate rotations. On the other hand, the fact that the rotation generator is a generator of the Lie algebra in the model guaranties that the generated excitation is indeed rotational invariant in the subspace. In addition we have the freedom to place our ``internal basis'' anywhere on the unit circle, where in our example in fig.~\ref{fig:gen3} an ``internal basis'' is shown on the $x_1$ and the $x_2$ axis. Further we know that quantum mechanical wave functions can be decomposed into eigenfunctions or eigenvectors which in the case of spin 1/2 are the $\ket{up}$ and $\ket{down}$ basis, for example, and we note that this decomposition is done in the internal space. One can follow this picture immediately in fig.~\ref{fig:gen3}, where we decompose an arbitrary vector in the two dimensional ``internal space'' into its two components. The two basis vectors for this decomposition commonly chosen, are the orthogonal eigenvectors of $g_1$, represented in fig.~\ref{fig:gen3}a and b as the small arrows on the horizontal $x_1$ axis, for example, pointing in $x_1$ direction in fig.~\ref{fig:gen3}a and in $x_2$ direction fig.~\ref{fig:gen3}b. The wave of eq.~\eqref{eq:gbasis} in Euclidean space used exactly this decomposition. Further according to Wigners \cite{wigner1939ireps} classification the \textit{little group} of a massless excitation the \textit{continuous spin} representation is the Euclidean group E(2) in two dimensions. The group E(2) splits into two translations and a rotation. This group is realized locally here. Again focusing onto the arrows on the $x_1$ axis, representing the change of the $x_1$ basis vector, we see in fig.~\ref{fig:gen3}a that the action of $g_1$ does the translation of the arrows tip in the horizontal $x_1$ direction, whereas the generator $g_2$ in fig.~\ref{fig:gen3}b does the translation in the vertical direction and finally $g_3$ the rotation. The last thing to understand from a geometric point of view is why we use the angle of $\frac{1}{2}\beta$ when rotations are done. First we note that we are looking to the transformation of the eigenvectors of $g_1$ applying rotations because in quantum mechanics the eigenvectors are the basis we use to represent a state. The matrix $g_1$ generates the twofold mirror symmetry of the symmetric ellipse. Therefore when rotating the deformation by an angle $\alpha$ in the subspace perpendicular to the propagation, the mirror symmetry of the deformation has to rotate by the same angle $\alpha$ but is described by using $\beta$:  $\alpha= \frac{1}{2} \beta$.

\section{Tensor rotations}
Surprisingly the gravitational waves of general relativity \cite{misner1973gravitation}, which are understood to be spin 2 waves, have the same space-time structure, with the + and x-polarization of gravitational waves corresponding to the $g_1$ and $g_2$ generators respectively. But in the previous section it was shown, that these waves belong to a irreducible spin 1/2 representation and the different spin would destroy the uniqueness of the angular momentum representations or alternatively the fermion commutator relations are without any meaning. Therefore a different solution is expected. It is well known that a symmetric tensor representation is reducible and can be decomposed into different spins \cite{cesare2017fieldapprtogw}. The general idea here is to relate these different angular momentum or spin representations of tensors to different representations used to represent the rotation of tensors. However we will focus on faithful representations only and in particular on the matrix representation and the representation of a tensor by vectors.
Starting with the vector representation we observe that we may represent the symmetric deformation tensor by an ellipsoid that is given by its semi-major axes $\vect{h_i}$, i=1..3 in case of the 3 dimensional space. Here we understand the ellipsoid as the mapping of the unit sphere. Then the semi-major axes of the ellipsoid correspond to the eigenvectors of the tensor, where the length of the semi-major axes is given by the corresponding eigenvalue. An explicit simplified example is given in eq.\eqref{eq:hirot}, where a tensor is represented by the semi-major axes $\vect{h_i}$ with length of $a_1,a_2$ and $a_3$ respectively. It will be shown that the length of the $\vect{h_i}$ are the eigenvalues of the matrix representation as well.
The semi-major $\vect{h_i}$ axes as vectors allow a coordinate invariant representation of the ellipsoid and therefore of the tensor and consequently the symmetric tensor and the three orthogonal semi-major axes have the same number of degrees of freedom. Further there is a unique bijective map $\xi:\{\vect{h_i}\}\mapsto \tens{A}$, where $\tens{A}$ is the matrix representation and $\xi$ maps the semi-major axes representation of the tensor to the matrix representation $\tens{A}$ of the tensor. The $\tens{A}$ is given as the unique solution of the linear eq.\eqref{eq:himap}, stating the properties of the semi-major axes and where the semi-major axes $\vect{h_i}$ are given. The inverse map $\xi^{-1}$ is computed by the same equation but solving for the unknown $\vect{h_i}$, which effectively is the calculation of the eigenvectors and eigenvalues of $\tens{A}$ and the $\vect{h_i}$ are the particular eigenvectors representing the semi-major axes.
\begin{equation} \label{eq:himap}
 \tens{A} \frac{\vect{h_i}}{|\vect{h_i}|} = \vect{h_i}; \quad\tx{$|\vect{h_i}|\neq0$; i=1..3}
\end{equation}
Geometrically the $\vect{h_i}$ are ordinary vectors and therefore a rotation of the ellipsoid or tensor can be done by a rotation of the semi-major axes, as shown in eq.\eqref{eq:hirot} where a rotation about the $x_3 $ axes is applied. Therefore the rotated tensor is represented by the rotated $\vect{h_i}(\alpha)$. Based on the rotated $\vect{h_i}(\alpha)$ one can solve eq.\eqref{eq:himap} to recover the matrix representation of the tensor. The resulting matrix $\tens{A}$ is given in eq.\eqref{eq:hirot} as well and is the same matrix representation as given by the standard matrix rotation in eq.\eqref{eq:rot}. It is obvious from the representation of the tensor by it's semi-major axes, that a rotation angle $\alpha=2\pi$ (see fig.~\ref{fig:gen3}c) gives a full turn of the ellipsoid corresponding to a spin 1 transformation. It follows that the semi-major axes representation is a faithful representation of a deformation tensor. However, when a symmetric tensor is represented by a symmetric matrix and for example the same rotation about the $x_3$ axis by an angle $\alpha$ is calculated
as in eq.\eqref{eq:rot}, where 
a diagonal matrix with eigenvalues of $a_1,a_2$ and $a_3$ is rotated about the $x_3$ axis, then $\alpha$ refers to the angle by which a vector rotation would be parametrized.
By recognizing that the angle $\alpha$ is multiplied by 2 one concludes that we have a spin 2 representation because the tensor does somehow two turns for a $2\pi$ angle of $\alpha$. However we know now this is not the case, if tensor means ellipsoid, and the rotation is parametrized by an angle from a semi-major axis to a reference axis.
\revision{\begin {widetext}
\begin{equation}\label{eq:hirot}
\begin{split}
&\vect{h_1}= \begin{pmatrix}
 a_1  \\
 0     \\
 0   
  \end{pmatrix},
\vect{h_2}= \begin{pmatrix}
 0  \\
 a_2 \\
 0   
  \end{pmatrix},
\vect{h_3}= \begin{pmatrix}
 0  \\
 0   \\
 a_3   
  \end{pmatrix},
  D_{x_3}(\alpha)=
  \begin{pmatrix}
 \cos(\alpha)  &-\sin(\alpha)    &0  \\
 \sin(\alpha)     &\cos(\alpha) &0  \\
 0      &0    &1 
  \end{pmatrix},\vect{h_i}(\alpha)= D_{x_3}(\alpha) \vect{h_i},\\
&\vect{h_1}(\alpha)=\begin{pmatrix}
 a_1\cos(\alpha)  \\
 a_1\sin(\alpha)  \\
 0   
  \end{pmatrix},
 \vect{h_2}(\alpha)=\begin{pmatrix}
 a_2\cos(\alpha) \\
 a_2\sin(\alpha) \\
 0   
  \end{pmatrix},
   \vect{h_3}(\alpha)=\begin{pmatrix}
 0  \\
 0  \\
 a_3   
  \end{pmatrix},\\
&\tx{solving eq.\eqref{eq:himap}: }\tens{A}=\begin{pmatrix}
 \frac{a_1+a_2+\left(a_1-a_2\right)cos(2\alpha)}{2} & \frac{\left(a_1-a_2\right)sin(2\alpha)}{2}& 0\\
 \frac{\left(a_1-a_2\right)sin(2\alpha)}{2} &\frac{a_1+a_2-\left(a_1-a_2\right)cos(2\alpha)}{2}& 0\\
 0 &0& a_3   \end{pmatrix}
  \end{split}
\end{equation}
\end {widetext}
} 

\begin {widetext}
\begin{equation}\label{eq:rot}
D_{x_3}(\alpha) \begin{pmatrix}
 a_1   &0    &0  \\
 0      &a_2  &0  \\
 0      &0    &a_3 
  \end{pmatrix}D^{-1}_{x_3}(\alpha)=
  \begin{pmatrix}
 \frac{a_1+a_2}{2}  &0    &0  \\
 0     &\frac{a_1+a_2}{2} &0  \\
 0      &0    &a_3 
  \end{pmatrix} +
 \begin{pmatrix}
 \frac{\left(a_1-a_2\right)}{2}\cos(2\alpha) &\frac{\left(a_1-a_2\right)}{2}\sin(2\alpha)     &0  \\
 \frac{\left(a_1-a_2\right)}{2}\sin(2\alpha)     &\frac{-\left(a_1-a_2\right)}{2} \cos(2\alpha) &0  \\
 0      &0    &0 
  \end{pmatrix}  
\end{equation}
\end {widetext}

But then the obvious question is: what
rotates two full turns for a single turn rotation of the ellipsoid? The answer is already given in the previous section where we have seen that the ``internal vectors'' rotate by $4\pi$ for a $2 \pi$ rotation of the ellipsoid. This is easily recognized when the matrix representation of the tensor is separated into the part unaffected by the rotation and the rotating part as in eq.\eqref{eq:rot}. Therefore the representation of a rotation with the matrix representation does so by rotating the ``internal vectors'' and has to multiply the angle $\alpha$ that is given as the parameter to describe the rotation by a factor of 2 in order to translate from the rotation angle $\alpha$ for vectors to the rotation angle  $\beta$ for the ``internal vectors''. The mirror symmetry which allows to change the sign of the eigenvectors without affecting the tensor reflects the fact that the ``internal vectors'' represent a double cover. The complementary representation of rotations is used by the spin 1/2 based representation, where the input parameter is the angle of the ``internal vectors'' $\beta$ describing the rotation of the ellipsoid as seen from a coordinate point of view. These different representations of rotations of the symmetric tensor are obvious when the tensor is represented as an ellipse as in fig.~\ref{fig:gen3}c and one can see how the different representations just encode the same rotation of the ellipse. Therefore it seems that the fact that a symmetric tensor admits a reducible representation of angular momentum is related to the different representations of rotations of the tensor and that the spin 1/2 representation uses a different reference angle compared to the standard matrix representation of tensors but both are complementary.

\section{Conclusion}
It has been shown that it is important to refer to the same representation of rotations when comparing spin. This is a problem for symmetric tensors where two different possible reference angles exist to describe the rotation of the tensor. Referring the rotation angle to the ``internal  space'' corresponds to the quantum mechanical picture, which is supported by the fact that the internal angular momentum of quantum theory is not related to coordinate rotations directly and by the presented construction of a deformation tensor that is a spin 1/2 wave in the quantum mechanical sense. It might be that this geometric point of view just holds for gravitational waves, which are derived by a geometric concept, and that it is impossible to expand it to particles physics more generally. But it suggests that a unified picture of gravitational waves of general relativity and the quantum theoretical approach to particles is possible. Independent thereof, we have a purely geometric picture of spin 1/2, which allows to teach the spin 1/2 concept on a more comprehensible basis. 

%

%
%



%
%
%
%

\end{document}